\documentstyle[12pt,epsfig,procsla]{article}
\catcode`\@=11
\long\def\@makefntext#1{
\protect\noindent \hbox to 3.2pt {\hskip-.9pt
$^{{\ninerm\@thefnmark}}$\hfil}#1\hfill}                

\def\@makefnmark{\hbox to 0pt{$^{\@thefnmark}$\hss}}  

\def\ps@myheadings{\let\@mkboth\@gobbletwo
\def\@oddhead{\hbox{}
\rightmark\hfil\ninerm\thepage}
\def\@oddfoot{}\def\@evenhead{\ninerm\thepage\hfil
\leftmark\hbox{}}\def\@evenfoot{}
\def\sectionmark##1{}\def\subsectionmark##1{}}

\newcommand{\dedx}  {\mathrm{d}E/\mathrm{d}x}
\newcommand{\epem}  {\mathrm{e}^+\mathrm{e}^-}

\newcommand{\zz}  {\mathrm{Z}}
\newcommand{\sr}  {\small\rm}
\newcommand{\unit}[1]{\,\mathrm{#1}}
\begin{document}
\thispagestyle{empty}
\centerline{\large\bf Multiparticle Aspects of $\epem$ Interactions at an}
\baselineskip=16pt
\centerline{\large\bf Energy of 133 GeV at LEP \footnote[1]{work 
supported by BMBF (Bundesministerium f\"ur Bildung, Wissenschaft, Forschung und 
Technologie)}}
\vspace{0.6cm}
\centerline{\normalsize Claus Grupen}
\baselineskip=13pt
\centerline{\footnotesize\it Department of Physics, University of Siegen}
\baselineskip=12pt
\centerline{\footnotesize\it D-57068 Siegen, Germany}
\baselineskip=12pt
\centerline{\footnotesize e-mail: claus.grupen@cern.ch}
\vspace*{0.9cm}
\enlargethispage{25mm}
\abstracts{Multiparticle aspects in $\epem$ collisions at center-of-mass
energies beyond the $\zz$ mass are investigated. To this end, data taken at 
130 and $136\,$GeV with the four LEP detectors ALEPH, DELPHI, L3 and OPAL with
an integrated luminosity of about $5 \unit{pb}^{-1}$ per experiment are
analyzed. A large number of events originate from radiative returns to the Z, 
i.e.~~$\epem \rightarrow \gamma\ \zz$ with well understood Z-decays. The
inter\-esting events with a visible energy corresponding to the c.m.s.~energy 
show
features as expected from the extrapolation from the Z. One experiment (ALEPH)
finds in four-jet final states some evidence for a clustering of the sum of
di-jet masses around $105\,$GeV, which however, is not supported by the three
other LEP collaborations. A preliminary analysis of data taken at a 
center-of-mass
energy of $\sqrt{s} = 161\,$GeV shows agreement with QCD expectations.}

\normalsize\baselineskip=15pt
\setcounter{footnote}{0}
\renewcommand{\thefootnote}{\alph{footnote}}

\section{Introduction}
The  large electron-positron collider (LEP) at CERN was upgraded in November
1995 from a center-of-mass energy of $\sqrt{s} = 91\,$GeV (on the Z peak) to
$\sqrt{s} = 130-140\,$GeV. The total number of events collected per experiment
was about 1500 corresponding to an integrated luminosity of 
$5.5 \unit{pb}^{-1}$. The bulk of data was taken at $130\,$GeV and $136\,$GeV
in about equal proportion. Only $0.04 \unit{pb}^{-1}$ was obtained at 
$140\,$GeV.

The running at an intermediate energy between LEP I (on the Z-peak) and LEP II
($\sqrt{s} \ge 2 \unit{M}_{\rm{W}}$) has been named LEP 1.5. LEP 1.5 can
check the running of $\alpha_{\rm{s}}$, the coupling constant of strong
interactions. Also, the measurement of infrared finite quantities like jet rates
and event shape distributions allows a comparison with perturbative QCD
calculations. Infrared sensitive quantities like multiplicities and inclusive
particle production cannot be calculated perturbatively, but the experimental
results can be compared to the energy evolution, which is predicted by
quantumchromodynamics (QCD). Possible deviations would hint at deficiencies of
particular Monte Carlo models or the presence of unexpected physical processes.

A drawback of LEP 1.5 is that the event rates are much lower compared to running
on the Z-peak. On the other hand the effects of hadronization, which cannot be
calculated in QCD, are predicted to be smaller at high $\sqrt{s}$. Therefore
perturbative 
\vspace{3mm}
\large
\begin{center}
\hspace{-5mm}
{\tt Invited talk given at the "XXVI International Symposium on
Multiparticle Dynamics" held in Faro, Portugal,\\ September 1-5, 1996}
\vspace{2mm}
\end{center}
\normalsize
QCD predictions  are expected to give a better description of the
data.

\begin{figure}[hbt]
\vspace{-0.5cm}
\begin{center}
\epsfig{file=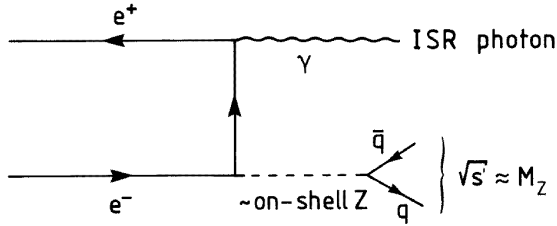, width=80mm}
\end{center}
\vspace{-0.5cm}
\caption{Illustration of radiative returns to the Z in
 $\epem$-interactions at a center-of-mass energy larger than $M_{\rm{Z}}$ .}
\label{fig1}
\end{figure}

One particular problem at LEP 1.5 is that the majority of events is radiative 
(Fig.~\ref{fig1}). Frequently the incoming electron or positron emits an 
energetic 
photon (usually at low polar angles) leading to an effective center-of-mass
energy close to the Z mass. These radiative returns to the Z are characterized
by two acollinear jets with an energetic photon, either detected in the forward 
calorimeters or escaped along the beam pipe.

The goal, of course, here is to study multiparticle dynamics at 
$\sqrt{s} = 130-140\,$GeV, and not to investigate the decay of almost real Z's.

To select a sample of hadronic  events  at the full center-of-mass energy one
can  tag initial state radiation photons with the luminosity  calorimeters or
use the event kinematics to reject radiative events. The separation between
radiative and non-radiative events is essential.

For physics comparisons the data taken at $130\,$GeV  and $136\,$GeV are energy
corrected to $\sqrt{s} = 133\,$GeV and averaged. Table 1 shows the number of
hadronic events obtained by the four LEP detectors. 

\begin{table}[hb]
\begin{center}
\begin{tabular}{|c|c|c|c|}
\hline
 {\sr ALEPH} & {\sr DELPHI} & {\sr L3} & {\sr OPAL} \\
\hline
\hline
 {\sr 299} & {\sr 346} & {\sr 402} & {\sr 291}\\ 
\hline
\end{tabular}
\caption[]{\sr Number of hadronic events at $\sqrt{s} = 133\,$GeV 
\cite{aleph1,delphi1,l31,opal1}.}
\label{numbers}
\end{center}
\end{table} 

The different event numbers are largely due to different cuts against radiative
events.

\section{Hadron Production}
Hadron Production in $\epem$-interactions proceeds in four steps \cite{grupen}:
\begin{enumerate}
\item The electroweak theory  describes the production of a gauge boson (Z or 
$\gamma$) in $\epem$-collisions.
\item The gauge boson produces a quark-antiquark pair which initiates a
quark-gluon cascade. As long as the energies of the partons are sufficiently
high this cascade can be described by perturbative QCD.
\item The transformation of coloured partons to colour-neutral hadrons
("hadronization") cannot be calculated by quantumchromodynamics but requires
Monte Carlo models which have been tuned to fit the data at 
$\sqrt{s} = 91\,$GeV.
\item Finally the end products of the hadronization process (primary hadrons,
colour-neutral clusters or "clans") decay into the observed final state hadrons.
Only longlived hadrons (lifetime $\ge 1\,$ns) are directly  seen in the
detectors.
\end{enumerate}  

When the data are compared to QCD predictions various hadronization models are
used. The JETSET event generator is based on the string fragmentation model. It
exists in a matrix element (up to order $\alpha_{\rm s}^{2}$) and a parton 
shower
version. It has been recently integrated into PYTHIA, which describes in additon
to $\epem$-interactions also hard ep an pp scattering. For $\epem$-annihilation
JETSET.PS and PHYTIA are equivalent. HERWIG is based on the cluster
fragmentation model, while ARIADNE uses radiation from colour dipoles with
string fragmentation. COJETS incorporates independent fragmentation. 

\section{Particle Identification}
\begin{figure}[htb]
\unitlength1cm
\begin{center}
{\epsfig{file=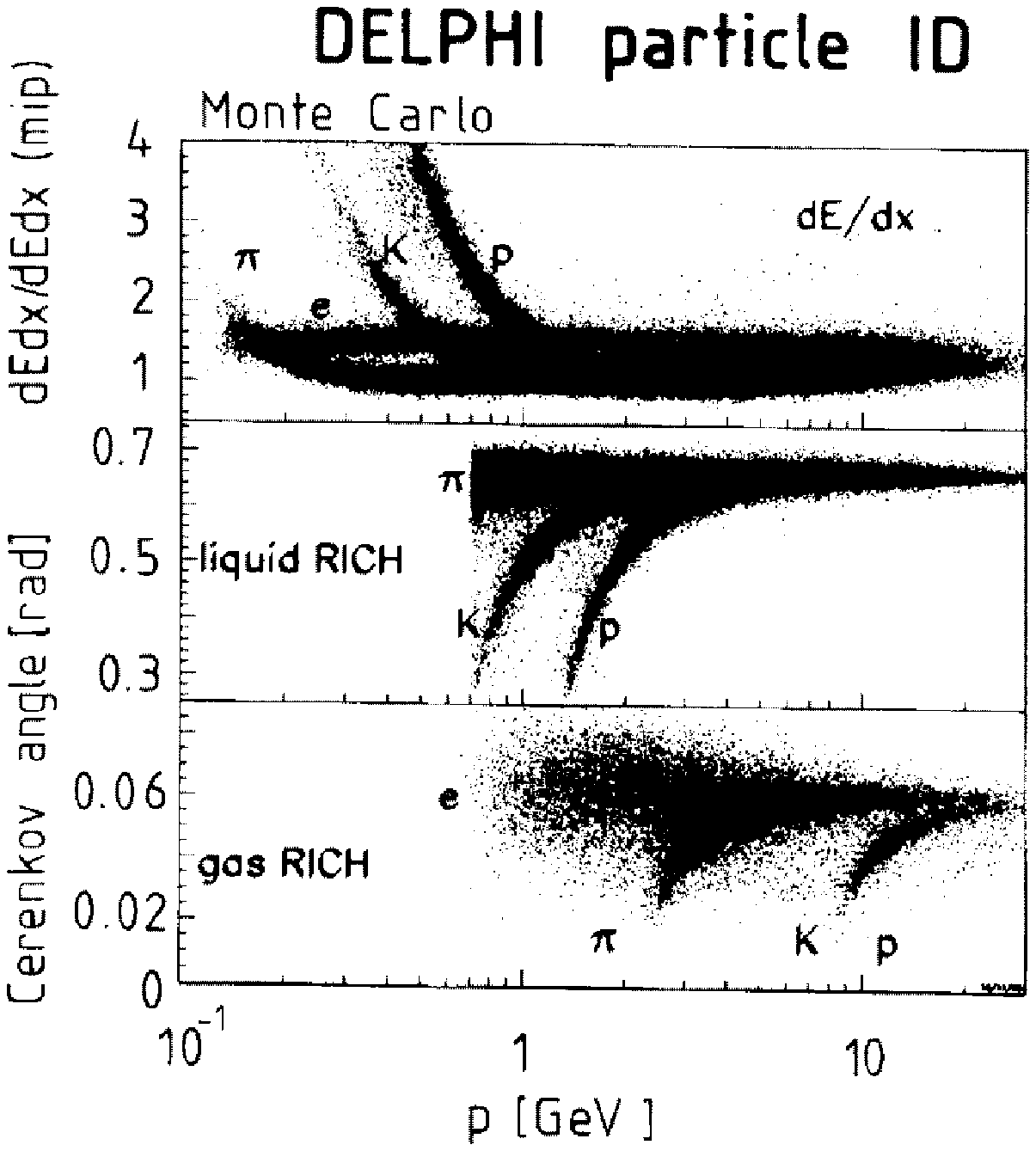,width=94mm}}
\vspace{-0.5cm}
\end{center}
\caption{$\dedx$ and RICH information for simulated
 charged particles in multihadronic events \protect\cite{delphi2}.}
\label{fig2a}
\begin{center}
{\epsfig{file=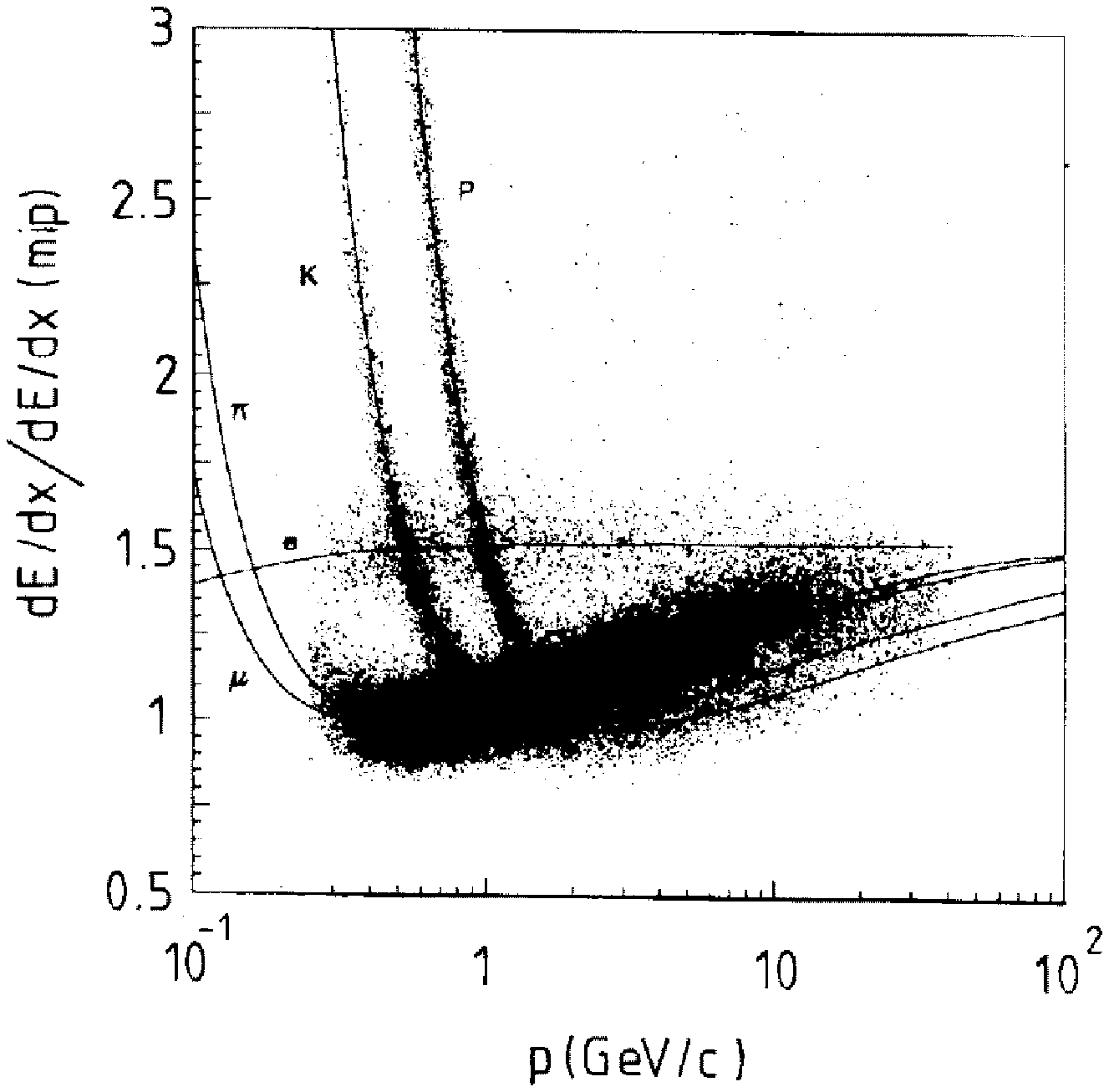,width=78mm}}
{\epsfig{file=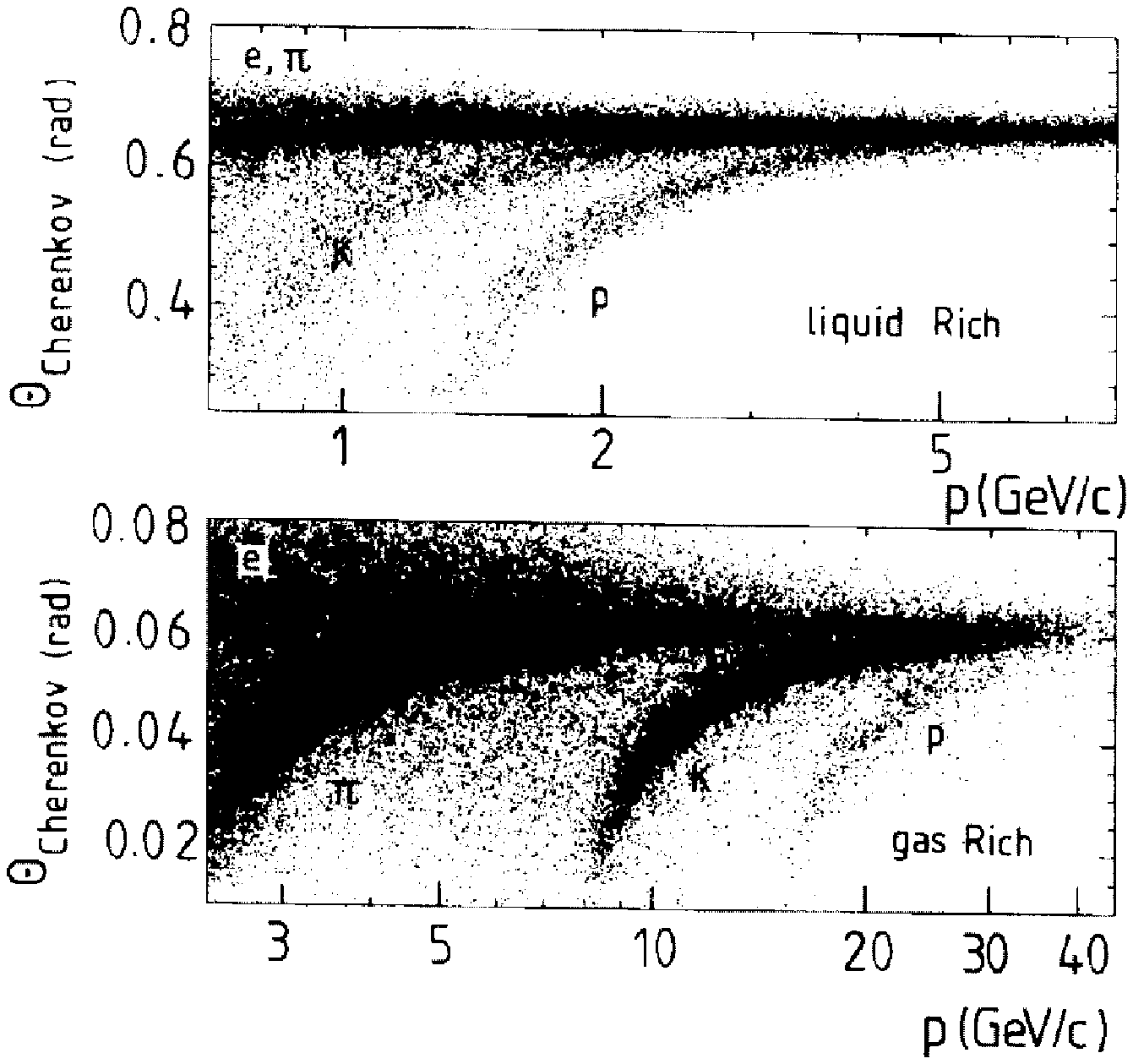,width=70mm}}
\vspace{-0.5cm}
\end{center}
\caption{$\dedx$ and RICH information for reconstructed charged particles in
 multihadronic events~\protect\cite{delphi2}.}
\label{fig2b}
\end{figure}

Particle identification in the four LEP detectors relies on the multiple
measurement of the specific energy loss $\dedx$ and/or the determination of
Cherenkov-ring radii in RICH counters. Fig.~\ref{fig2a} shows the $\dedx$ and
RICH-information for a set simulated and Fig.~\ref{fig2b} for observed events
\cite{delphi2}. 
 Clearly, a $\pi$-K-p separation can be obtained in certain kinematic ranges.
The characteristic shower profiles in electromagnetic and hadronic calorimeters 
can be used to distinguish electrons from hadrons, and the penetration through
hadron calorimeters identifies muons. For shortlived particles ($100\,$fs $\le
\tau \le$ $1\,$ns) secondary vertices can be reconstructed and very shortlived
particles are identified  by the invariant mass of their decay products.

\section{Total Cross-Section and Global Event Properties}
\begin{figure}[htb]
\unitlength1cm
\begin{center}
{\epsfig{file=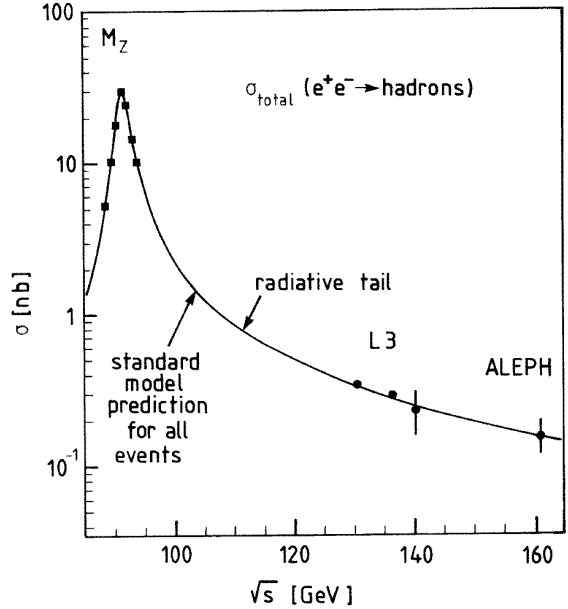,width=80mm}}
\vspace{-0.5cm}
\end{center}
\caption{Cross-section for the process $\epem \rightarrow \unit{hadrons}$.
 The solid line is the Standard Model prediction for the total cross section
 \protect\cite{l32}.}
\label{fig3}
\end{figure}

The total cross-section $\sigma_{\rm{tot}}$ ($\epem \rightarrow \unit{hadrons}$)
is well described by the Standard Model. The cross-section is dominated by a
pronounced radiative tail. Fig.~\ref{fig3} \cite{l32} shows the results of the
L3
experiment with data points at 130, 136 and $140\,$GeV along with lower energy
data. Also shown is the result of the ALEPH experiment \cite{aleph2} from the
recent run at $\sqrt{s} = 161\,$GeV. If the data samples are restricted to 
final states with visible hadronic energies close to the respective
center-of-mass energy also agreement with expectation is observed.

\begin{figure}[htb]
\unitlength1cm
\begin{center}
{\epsfig{file=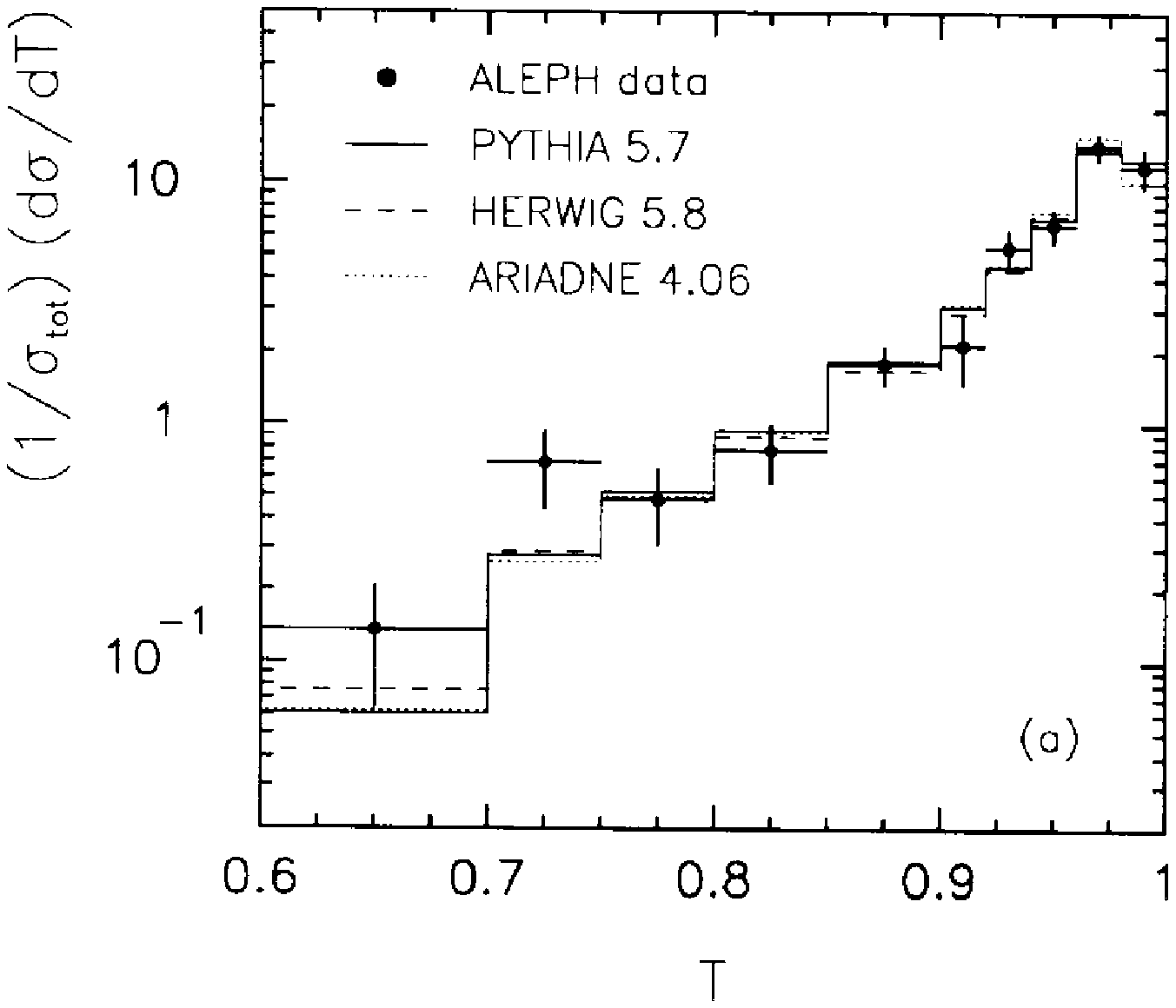,width=72mm}}
{\epsfig{file=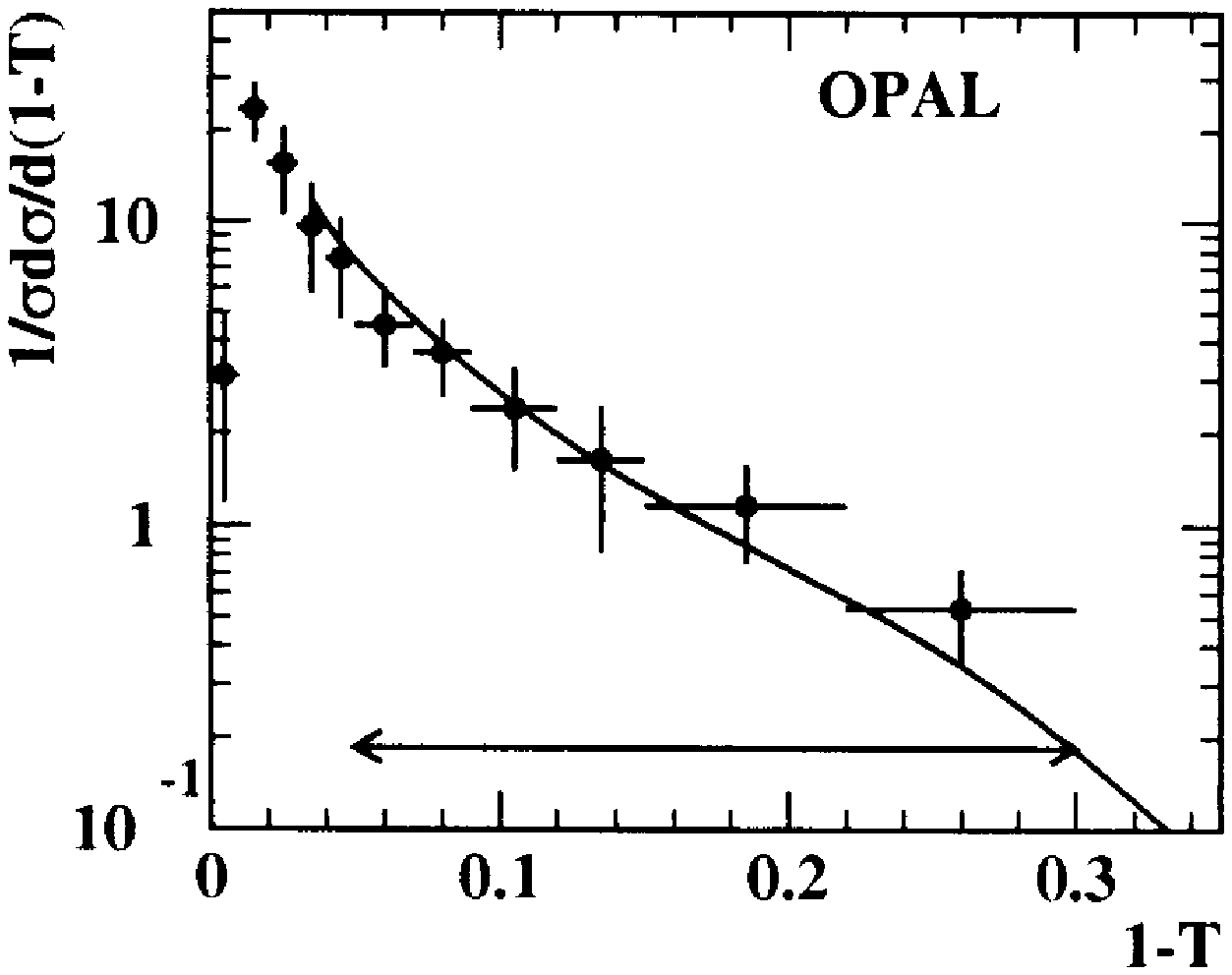,width=78mm}}
\vspace{-1cm}
\end{center}
\caption{Thrust distribution at a center-of-mass energy of 133$\,$GeV 
\protect\cite{aleph1,opal1}.}
\label{fig4}
\end{figure}

\begin{figure}[htb]
\unitlength1cm
\begin{center}
{\epsfig{file=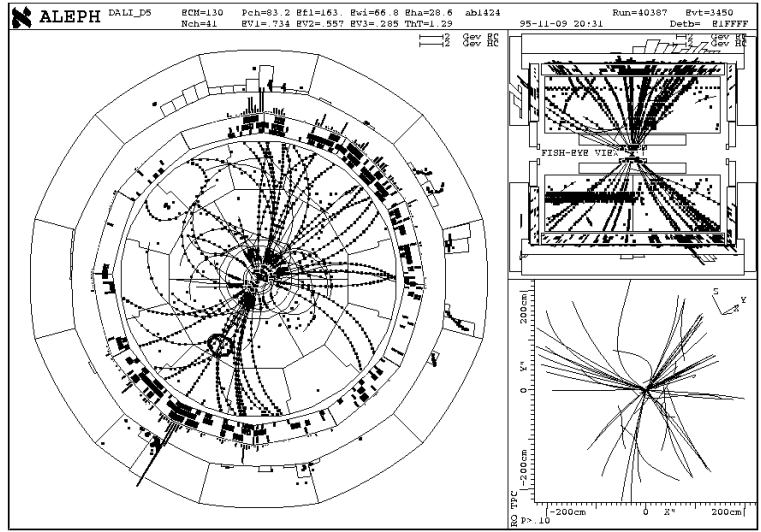,width=100mm}}
\end{center}
\caption{ALEPH event display of a low thrust, high multiplicity event
\protect\cite{aleph3}.}
\label{fig5}
\end{figure}

The global event shapes of hadronic final states can be characterized by
topological variables. The thrust-distibutions (Fig.~\ref{fig4}) observed by 
ALEPH \cite{aleph1} and OPAL \cite{opal1}, where
\begin{displaymath}
 T = \max_{\vec n} \left( \frac{\displaystyle{\sum_{i}} \mid \vec p_{i} \cdot 
 \vec n \mid}{\displaystyle{\sum_{i}} \mid \vec p_{i} \mid} \right) 
\end{displaymath}
($\vec p_{i}$ are the momenta of the final state hadrons), show in general good
agreement with Monte Carlo models, maybe with the exception of a small excess 
at low thrust values observed by ALEPH. 
Fig.~\ref{fig5} \cite{aleph3} shows as an example an event display of such a 
low thrust,
near spherical event. The energy evolution of $\langle T \rangle$ is well 
described by the models.

\begin{figure}[htb]
\unitlength1cm
\begin{center}
{\epsfig{file=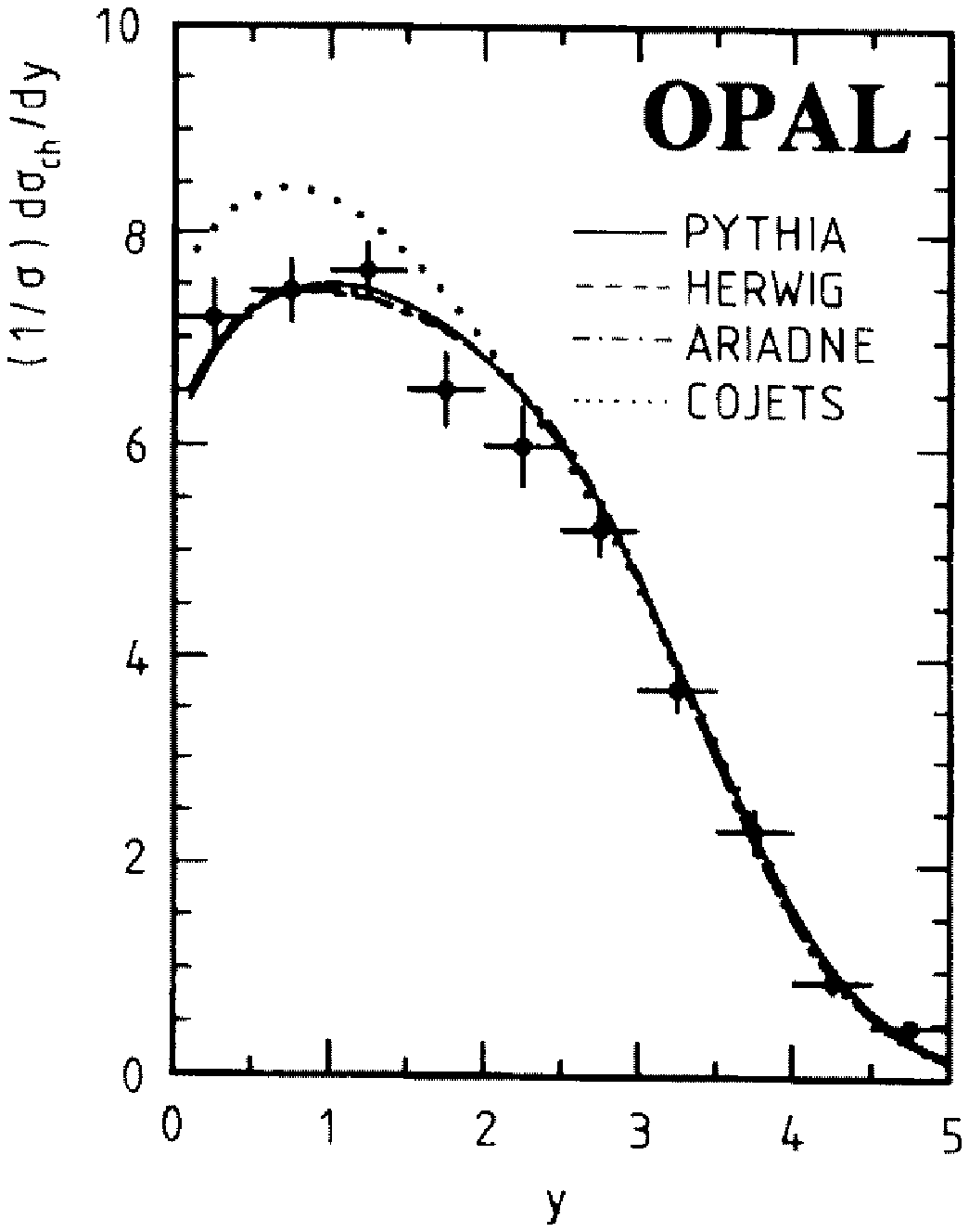,width=64mm}}
{\epsfig{file=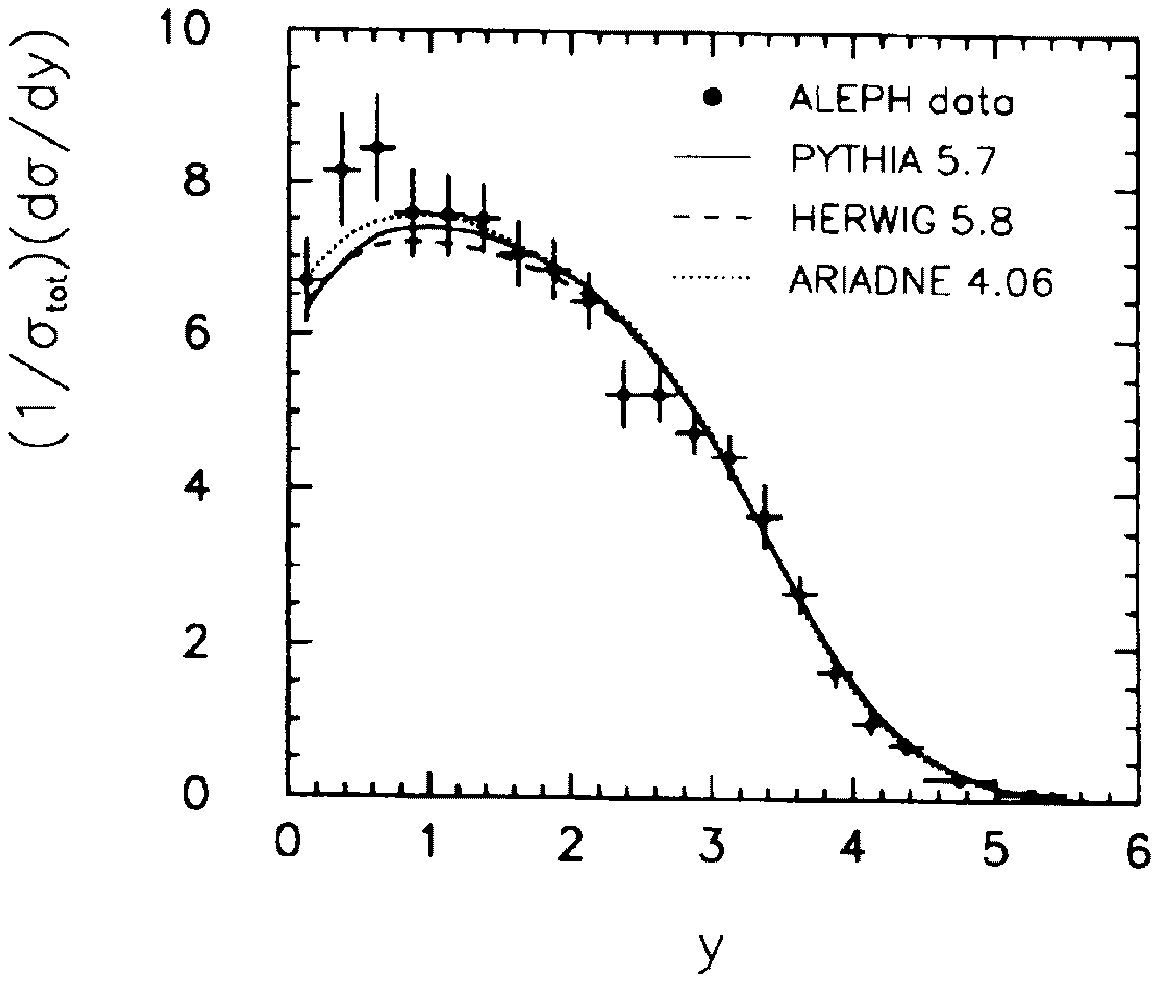,width=84mm}}
\vspace{-0.5cm}
\end{center}
\caption{Rapidity distribution from ALEPH \protect\cite{aleph1} and OPAL
\protect\cite{opal1}.}
\label{fig6}
\end{figure}

The distributions in rapidity
\begin{displaymath}
 y = \frac{1}{2} \left| \ln \frac{E+p_{\parallel}}{E-p_{\parallel}} \right|;
 \hspace{5mm} p_{\parallel} = \vec p \cdot \vec n_{\rm{thrust}}
\end{displaymath}
from ALEPH \cite{aleph1} and OPAL \cite{opal1} also show agreement with models
(Fig.~\ref{fig6} \cite{aleph1,opal1}). The small excess seen by ALEPH at low 
rapidities
is related to the excess of low thrust events. The OPAL data clearly indicate
that the COJETS Monte Carlo model with independent fragmentation fails to
describe the data at low rapidities. 

\begin{figure}[htb]
\unitlength1cm
\begin{center}
{\epsfig{file=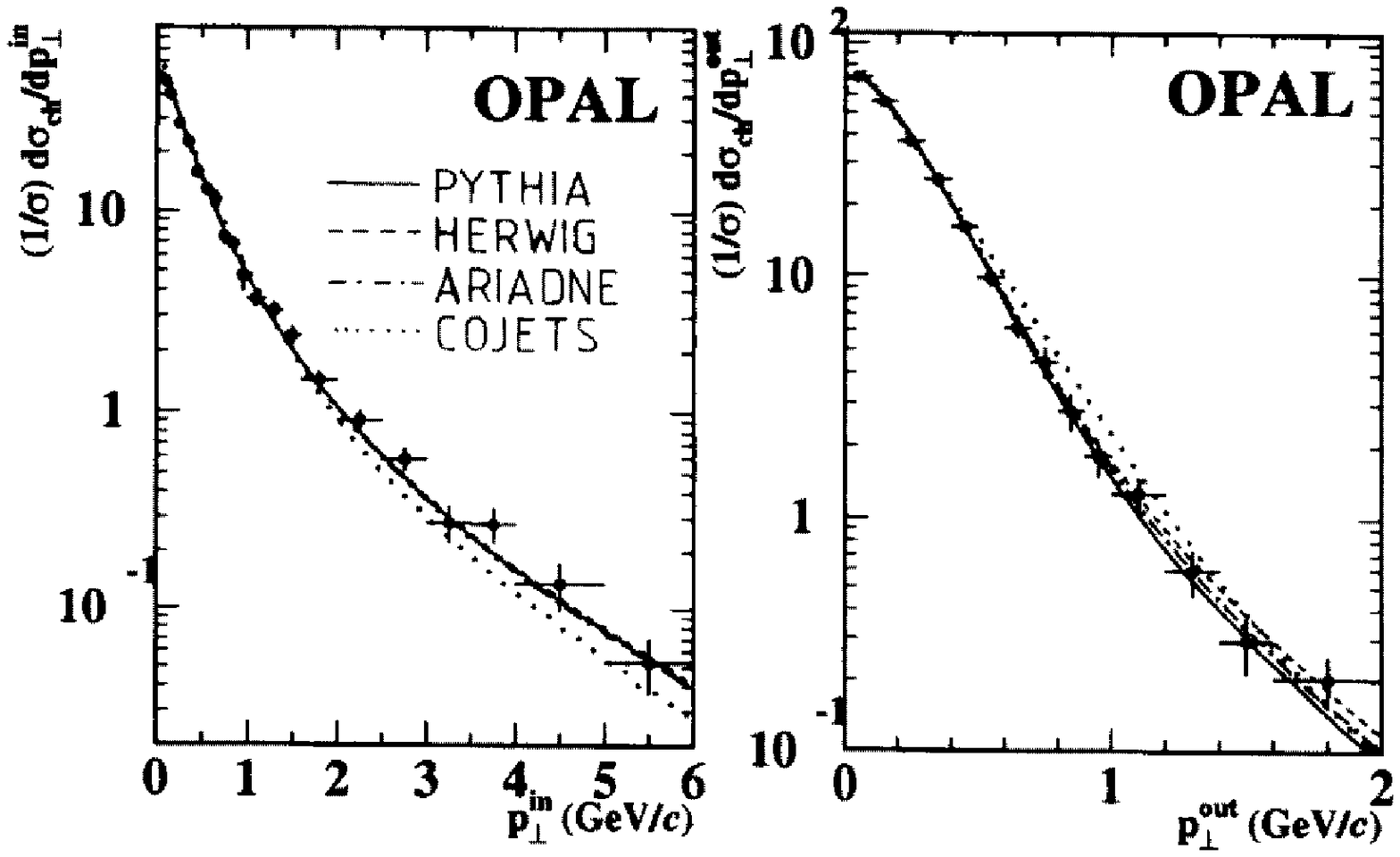,width=130mm}}
{\epsfig{file=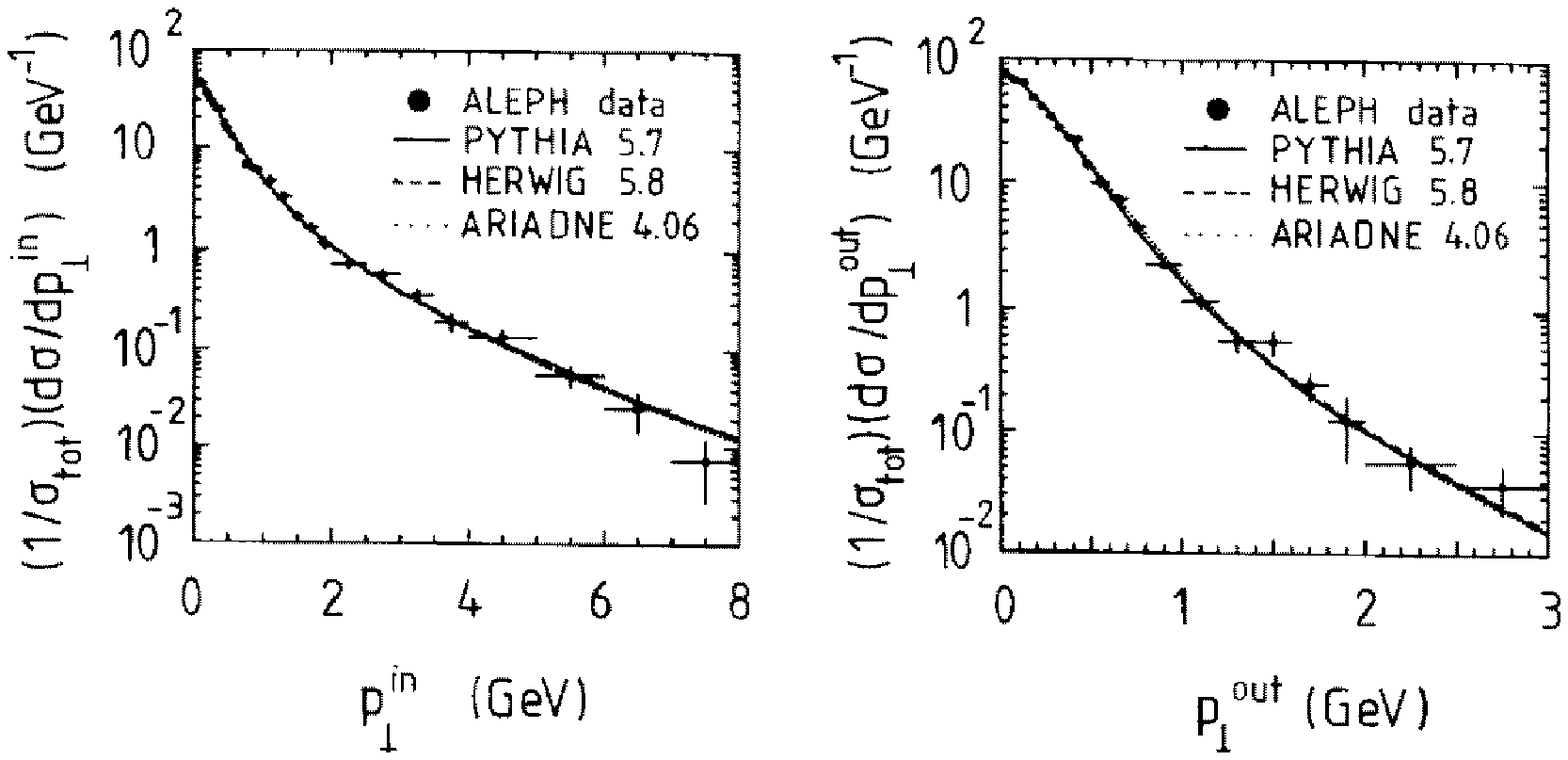,width=130mm}}
\vspace{-0.5cm}
\end{center}
\caption{Transverse momentum distribution 
 $p^{\rm{in}}_{\perp}$ and $p^{\rm{out}}_{\perp}$
 from ALEPH \protect\cite{aleph1} and \protect\newline 
 OPAL \protect\cite{opal1}.} 
\label{fig7}
\end{figure}

The sphericity tensor allows to investigate the planarity of events. It is given
by
\begin{displaymath}
 S^{\alpha \beta} = \frac{ \displaystyle{\sum_{i}} p_{i}^{\alpha} \hspace{1mm}
 p_{i}^{\beta}}{ \displaystyle{\sum_{i}} p_{i}^{2}};
\hspace{5mm} \alpha, \beta = 1, 2, 3
\end{displaymath}
where $p_{i}$ are the final state particle momenta with their components
$\alpha, \beta = 1, 2, 3$. This momentum tensor permits to define the transverse
momentum in the event plane $p^{\rm{in}}_{\perp}$ with respect to
the major sphericity axis and also the transverse momentum out of the event
plane $p^{\rm{out}}_{\perp}$. Within the statistics agreement
between observation and model predictions is observed (Fig.~\ref{fig7} 
\cite{aleph1,opal1}). The disagreement between experimental data and models 
observed at $\sqrt{s} = 91\,$GeV in the $p^{\rm{out}}_{\perp}$ distribution at
large $p^{\rm{out}}_{\perp}$ 
[e.g. \cite{rudolph}] is not confirmed at $\sqrt{s} = 133\,$GeV. Agreement is 
also observed at the recent high energy run at $\sqrt{s} = 161\,$GeV, albeit 
at low statistics.

The comparison of jet rates with prediction requires a jet counting algorithm.
Jets are normally defined by the Durham clustering scheme. The quantity
\begin{displaymath}
 y_{ij} = 2 \ \frac{\min \left(E_{i}^{2},E_{j}^{2} \right) \ \left(
 1 - \cos\, \theta_{ij} \right)}{E_{\rm{vis}}^{2}}
\end{displaymath} 
with $E_{i}, E_{j}$ - particle energies and $\theta_{ij}$ - opening angle
between the two particles, is calculated for all two-particle combinations. The
pair with the smallest $y_{ij}$ is replaced by a pseudoparticle by adding the
four momenta of the two particles. For a given $y$-value $y_{\rm{cut}}$ the
number of pseudoparticles gives the number of jets $N_{\rm{J}}$. If $y$ is
extremely small $N_{\rm{J}}$ is equal to the multiplicity of charged particles 
in the
event. $y_{\rm{cut}} \approx 0.1$ selects symmetric three-jet events, and for
even larger values the particles are clustered into two jets. Instead of the
Durham clustering scheme, also the JADE-algorithm can be used by defining
\begin{displaymath}
 y_{ij} = \frac{2\,E_{i}\,E_{j} \ \left(
 1 - \cos\, \theta_{ij} \right)}{E_{\rm{vis}}^{2}}
\end{displaymath}
with an analogous iteration procedure.

The ALEPH results on jet rates as a function of the jet resolution parameter
$y_{\rm{cut}}$ displayed in Fig.~9 \cite{aleph1} show in general good 
agreement
with model predictions, although JETSET ${\cal O}\left(\alpha_{\rm{s}}^{2} 
\right)$
fails to reproduce the 5-jet rate. This is not a surprise because the matrix
element version of JETSET ${\cal O} \left(\alpha_{\rm{s}}^{2} \right)$, which is
 only up to
second order in $\alpha_{\rm{s}}$, can produce only four partons ($\epem
\rightarrow \rm{q}\, \rm{\bar q}\, \rm{g}\, \rm{g} \quad \rm{or} \quad \epem 
\rightarrow \rm{q}\, \rm{\bar q}\, \rm{q}\, \rm{\bar q}$).
\begin{figure}[htb]
\unitlength1cm
\begin{center}
{\epsfig{file=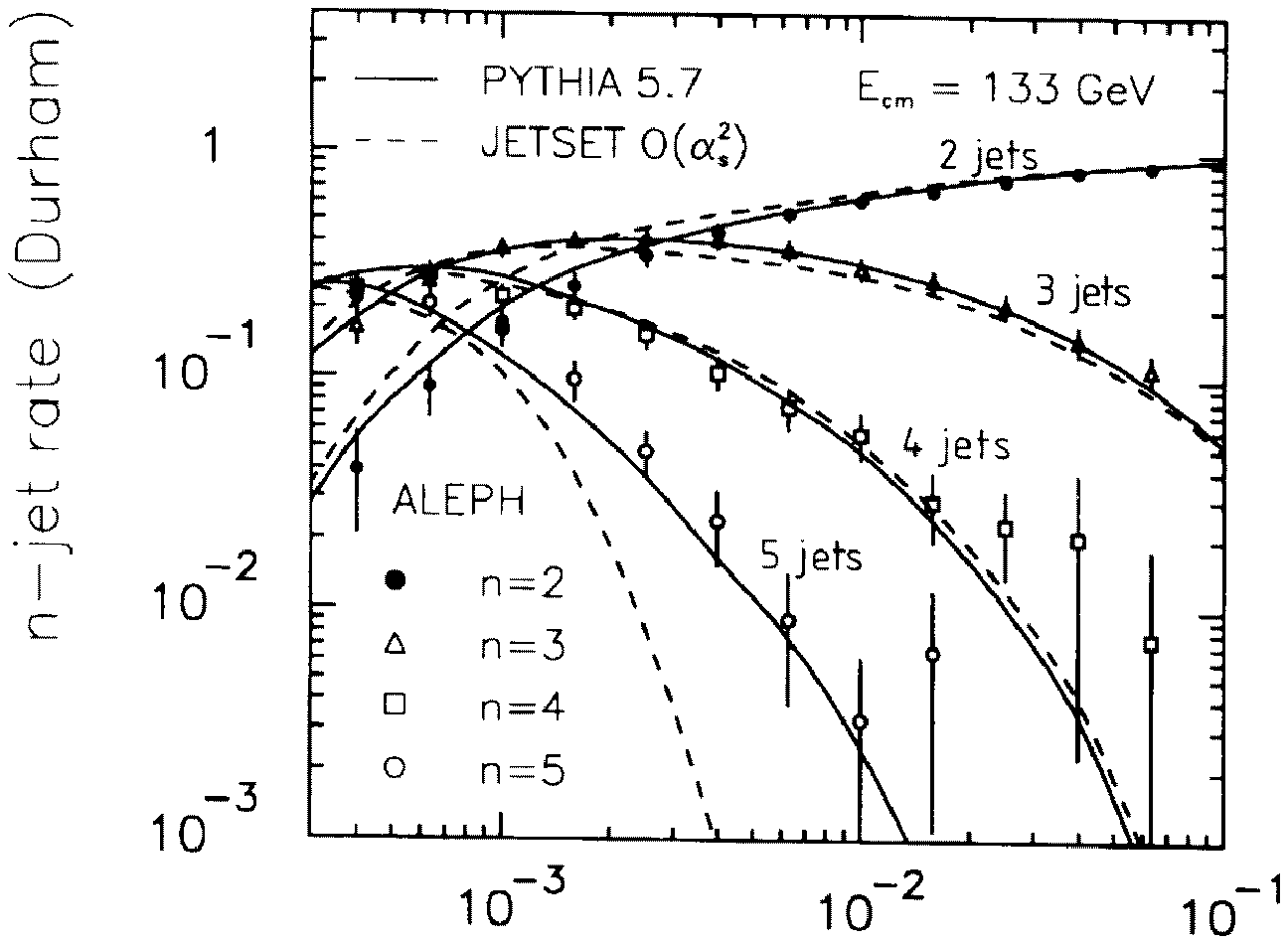,width=90mm}}
\vspace{-0.5cm}
\end{center}
\begin{center}
\caption{n-jet rates in their dependence on the $y_{\rm{cut}}$ value from 
ALEPH \protect\cite{aleph1} at $133\,$GeV.}
\end{center}
\label{fig8}
\end{figure}

The $\alpha_{\rm{s}}$-values derived from the global event properties of the 
four
LEP experiments show agreement among each other leading to a LEP average of
\begin{displaymath}
 \alpha_{\rm{s}} = 0.110 \pm 0.006 \hspace{5mm} \rm{at} \quad \sqrt{s} = 
 133 \unit{GeV}.
\end{displaymath}
Compared to the LEP I value of \cite{schmelling}
\begin{displaymath}
 \alpha_{\rm{s}} = 0.118 \pm 0.005
\end{displaymath}
the LEP 1.5 result ist lower by about one standard deviation, confirming the
running of the strong coupling constant.

\section{Multiplicities}
The amount of data is sufficient to determine average  charged multiplicities at
130 and $136\,$GeV separately. To study multiplicity distributions in detail
requires more data so that at the moment no firm conclusion can be drawn on
whether  KNO-scaling is valid or whether the distributions are best
reproduced  by a log-normal or negative binomial distribution.

\begin{figure}[htb]
\unitlength1cm
\begin{center}
{\epsfig{file=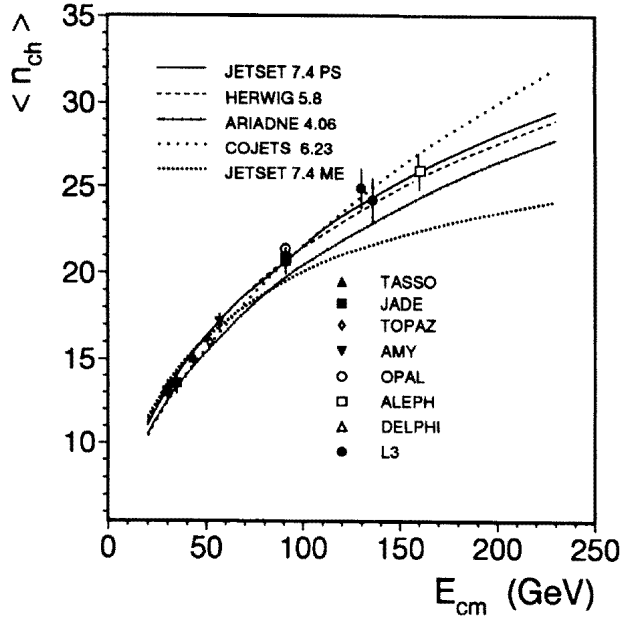,width=90mm}}
\vspace{-0.5cm}
\end{center}
\caption{Average charged multiplicity as a function of the center-of-mass energy
 \protect\cite{l31}.}
\label{fig9}
\end{figure}

The results of the four LEP experiments on the average charged multiplicity in
comparison to model predictions are shown in Table \ref{table2}  
\cite{aleph1,delphi1,l31,opal1}. Fig.~\ref{fig9} shows L3 data in comparison to 
results
at lower center-of-mass energies \cite{l31}. Also the average charged
multiplicity from the recent run at $161\,$GeV obtained by ALEPH, $\langle 
n_{\rm{ch}} \rangle \displaystyle{\mid_{161 \unit{GeV}}} = 25.8 \pm 1.0$ is
shown in the figure.

\begin{table}[hb]
\begin{center}
\begin{tabular}{|c|c|c|c|}
\hline
 { } & {\sr 130 GeV} & {\sr 133 GeV} & {\sr 136 GeV} \\
\hline
\hline
 {\sr ALEPH} & {\sr $23.61 \pm 0.65$} & {\sr $24.15 \pm 0.55$} & {\sr $25.01 \pm
 0.79$}\\
\hline
 {\sr DELPHI} & {\sr $23.84 \pm 0.73$} & {\sr } & {\sr }\\
\hline
 {\sr L3} & {\sr $24.9 \pm 0.9$} & {\sr } & {\sr $24.2 \pm 1.1$}\\
\hline
 {\sr OPAL} & {\sr } & {\sr $ 23.40 \pm 0.65$} & {\sr } \\
\hline  
\hline
 {\sr JETSET.PS} & { } & {\sr 24.2} & { }\\
\hline
 {\sr HERWIG} & { } & {\sr 24.1} & { }\\
\hline
 {\sr ARIADNE} & { } & {\sr 24.1} & { }\\
\hline
 {\sr JETSET ${\cal O}\left( \alpha_{\rm{s}}^{2} \right)$} & { } & {\sr 22.6} & 
 { }\\
\hline
\hline
\end{tabular}
\caption[]{\sr LEP results on the average charged multiplicity $\langle  
 n_{\rm{ch}} \rangle$ in comparison with model predictions
\cite{aleph1,delphi1,l31,opal1}.}
\label{table2}
\end{center}
\end{table} 

The experimentally observed energy evolution of the charged particle 
multiplicity
practically rules out the matrix element version of JETSET ${\cal O}\left(
\alpha_{\rm{s}}^{2} \right)$ and the COJETS model with independent 
fragmentation. JETSET.PS and HERWIG  describe the data rather well, while 
ARIADNE predicts somewhat low values.

\section{Inclusive Particle Production}
\begin{figure}[htb]
\unitlength1cm
\begin{center}
{\epsfig{file=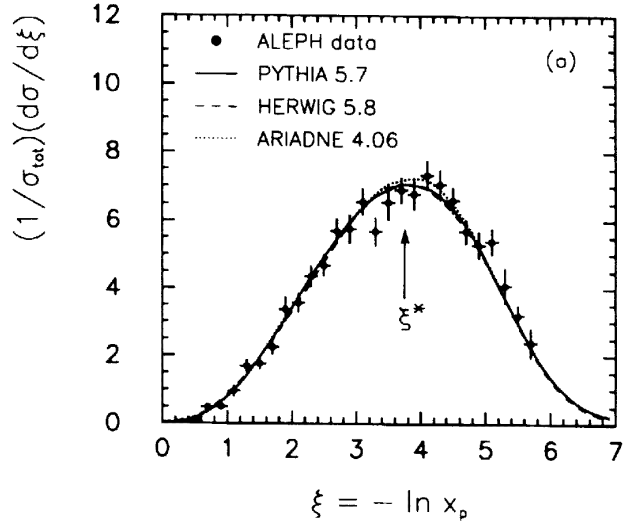,width=90mm}}
\vspace{-0.5cm}
\end{center}
\caption{All particle $\xi$ distribution at $133\,$GeV observed by ALEPH
.}
\label{fig10}
\end{figure}

The small amount of data does not allow a meaningful determination of inclusive
particle production rates for pions, kaons and protons separately. 
Fig.~\ref{fig10}
\cite{aleph1} shows the all particle momentum distribution in terms of the
variable $\xi = \ln \, 1/x_{p}$, where $x_{p} = p_{\rm{hadron}}/p_{\rm{beam}}$.
The variable $\xi$ emphasizes the low momentum region (large $\xi$ values).
Good agreement is observed with the Monte Carlo models. Also the energy
evolution of the maximum of the $\xi$-distribution $\left( \xi^{\ast} \right)$
is well described by a modified leading-log approximation (MLLA) \cite{opal1}.

\section{Four Jet Events}
In a search for $\epem \rightarrow \rm{X_{1}} \, \rm{X_{2}} \rightarrow 
\rm{four \ jets}$, where $\rm{X_{1}}$ and $\rm{X_{2}}$ are heavy particles 
decaying into two jets each, ALEPH has found a few events with interesting 
features. If the hadronic final states are clustered into four jets, if a
rescaling of the jet energies using the well measured jet directions is made, 
and
if QCD events are suppressed by a cut on the di-jet masses, a clustering of nine
events in the sum of di-jet masses at $105\,$GeV is observed where the pairing 
is
made by grouping two jets together which lead to the smallest di-jet mass 
difference
\cite{aleph4}. The expected QCD background in the $6.3\,$GeV-wide mass window
 centered on the $105\,$GeV mass peak is low ($<$ 1 event).
An analysis of the other three LEP-collaborations does not confirm the ALEPH
finding, although DELPHI, L3 and OPAL in a similar analysis also find somewhat
more events than predicted, but without any sign of clustering at a particular 
mass \cite{rolandi}.
The few anomalous ALEPH events are also responsible for the minor excess
observed in the low thrust and low rapidity region (Fig.~\ref{fig4} and 
\ref{fig6}). 
The full
analysis of the measurements at  $\sqrt{s} = 161\,$GeV taken in July/August 
1996 and the
high energy run at the end of this year (presumably at $\sqrt{s} = 172\,$GeV) 
will probably
settle the question whether the ALEPH excess is real or just a statistical
fluctuation.

\section{Running at a center-of-mass energy of 161$\,$GeV}
The LEP experiments have collected about $11 \unit{pb}^{-1}$ each
at $\sqrt{s} = 161\,$GeV in a run in July/August 1996. All experiments have 
seen W$^{+}\,$W$^{-}$ -pair events ($\sim$ 30 events per experiment) in the 
expected final states ($\rm{q}\, \rm{\bar q}\, \rm{q}\, \rm{\bar q}\, ; \ 
\rm{q}\, \rm{\bar q}\, \rm{l}\, \nu\, ; \ \rm{l}\, \nu\, \rm{l}\, \nu$) with 
the predicted cross-sections.

A preliminary analysis of the QCD events shows good agreement with 
expectations. 

\section{Conclusions}
The analysis of the data taken at $\sqrt{s} = 133\,$GeV shows good agreement for
the average charged multiplicity, the maximum value of the $\xi$-distribution, 
$\xi^{\ast}$, and the strong coupling constant in comparison to expectations
from QCD. The global event properties are well described by JETSET.PS, HERWIG
and ARIADNE Monte Carlo models, while COJETS (with independent fragmentation)
and  JETSET ${\cal O}\left( \alpha_{\rm{s}}^{2} \right)$ are less successful. 
ALEPH has
found a small sample of potentially interesting four-jet events. The evidence,
however, is not supported by DELPHI, L3 and OPAL, and also not by a first look
at the data taken at $\sqrt{s} = 161\,$GeV.

\section{Acknowledgements}
The author is grateful to the conference organizers for the hospitality and the
pleasant atmosphere at the symposium. 
I also thank Claudia Hauke and Volker Schreiber for their help in preparing the
written version of my talk.



\section{References}
\begin{thebibliography}{99}
\bibitem{aleph1}
 ALEPH-Collaboration ~ D.~Buskulic et al.~CERN-PPE/{\bf 96-43} (1996)
\bibitem{delphi1}
 DELPHI-Collaboration ~ P.~Abreu et al.~CERN-PPE/{\bf 96-05} (1996) \newline
 and {\it Phys.~Lett} {\bf B372} (1996) 172
\bibitem{l31}
 L3-Collaboration ~ M.~Acciarri et al.~CERN-PPE/{\bf 95-192} (1995) \newline
 and {\it Phys.~Lett} {\bf B371} (1996) 137
\bibitem{opal1}
 OPAL-Collaboration ~ G.~Alexander et al.~CERN-PPE/{\bf 96-47} (1996)
\bibitem{grupen}
 C.~Grupen {\it Proc.~XXV Int.~Symp.~on Multiparticle Dynamics 1995} \newline
 Stara Lesna; ed.~D.~Bruncko, L.~\v S\'andor, J.~Urb\'an ;p.~395, and 
 references therein
\bibitem{delphi2}
DELPHI-Collaboration ~ P.~Abreu et al.~CERN-PPE/{\bf 95-194} (1995)
\bibitem{l32}
 L3-Collaboration ~ M.Acciarri et al.~CERN-PPE/{\bf 95-191} (1995)
\bibitem{aleph2}
 ALEPH-Collaboration ~ D.~Buskulic et al. {\it Contribution to the
 Int.~Conf.~on High Energy Physics,} Warsaw 1996
\bibitem{aleph3}
 ALEPH-Collaboration ~ D.~Buskulic et al. DALI-Program, H.~Drevermann (1995)
\bibitem{rudolph}
 G.~Rudolph, these proceedings
\bibitem{schmelling}
 M.~Schmelling {\it Phys.~Scripta} {\bf 51} (1995) 683
\bibitem{aleph4}
 ALEPH-Collaboration ~ D.~Buskulic et al.~CERN-PPE/{\bf 96-52} (1996) \newline
 and {\it Z.~Phys.} {\bf C71} (1996) 179
\bibitem{rolandi}
 G.~Rolandi, ALEPH-Physics Note 96-20 (1996)
\end{thebibliography}
\end{document}